\documentstyle[12pt,onecolumn]{elsart}
\begin{document}
\addtolength{\topmargin}{50pt}
\def\today{24 March 2000}


\def\gsim{\mathrel{%
   \rlap{\raise 0.511ex \hbox{$>$}}{\lower 0.511ex \hbox{$\sim$}}}}
\def\lsim{\mathrel{
   \rlap{\raise 0.511ex \hbox{$<$}}{\lower 0.511ex \hbox{$\sim$}}}}


\def\simeqq{\,\,{\raise 1.2ex\hbox{?}\llap{$\simeq$}}}

\def\+{\mathrel{
   \rlap{\raise -0.970ex \hbox{$\mathbf\widehat{}$}}
{\lower 0.511ex \hbox{$\mathbf\bigcirc$}}}}
\def\-{\,\,\,\,\,\mathrel{
   \llap{\raise -0.500ex \hbox{$\mathbf\bigcirc$}}
{\lower 0.990ex \hbox{$\mathbf\widehat{}$}}}}
\def\beq{\begin{eqnarray}}
\def\eeq{\end{eqnarray}}
\def\e{\mbox{e}}
\def\eqb{\begin{eqnarray}}
\def\eqe{\end{eqnarray}}




\begin{frontmatter}

\title{\sc Probing Quantum Aspects of Gravity\thanksref{C}}

\thanks[C]{Work supported by Consejo Nacional de Ciencia y 
Tecnolog\'ia (CONACyT) under grant number 32067-E.}

\author{G.\ Z.\ Adunas, E.\ Rodriguez-Milla, D.\ V.\ Ahluwalia} 
\address{Theoretical Physics Group, ISGBG, Ap. Pos. C-600\\
Escuela de Fisica de la UAZ, 
Zacatecas, ZAC 98068, Mexico.\\
E-mail: ahluwalia@phases.reduaz.mx; http://phases.reduaz.mx}


\begin{abstract}
We emphasize that a specific aspect of quantum gravity is the 
{\em absence} of a super-selection rule that prevents a linear 
superposition of different gravitational charges. As an immediate 
consequence, we obtain a tiny, but observable, violation of the 
equivalence principle, provided, inertial and gravitational masses 
are not assumed to be operationally identical objects. In this framework, 
the cosmic gravitational environment affects local experiments. 
A  range of terrestrial experiments,  from neutron interferometry to 
neutrino oscillations, can serve as  possible probes to study the 
emergent quantum aspects of gravity.
\end{abstract}

\end{frontmatter}



\section{Introduction}

It should not at all be surprising that despite the usual ``forty orders of 
magnitude argument'' against a feasible quantum gravity phenomenology  
certain quantum aspects of gravity can be probed terrestrially.
Just as the exceedingly small cross sections for neutrinos at 
accessible energies, and the anticipated life time of  proton 
which hugely exceeds the present  age of the universe, have not discouraged 
a robust, and a highly rewarding, program of high energy phenomenology; 
similarly, it is now emerging that the  mere smallness of the coupling 
associated with the gravitational interactions of elementary particles, 
or the smallness of the Planck length in comparison, say,  with the 
size of atomic nuclei, does not prevent an experimental quantum-gravity 
phenomenology program that probes quantum gravity from its low energy 
limit to all the way up to Planck scale \cite{gacN,dvaN}.
   
In fact, about a decade ago Audretsch, Hehl, and 
L\"ammerzahl \cite{ahl} argued as to why matter wave interferometry and
quantum objects are fundamental to establishing an empirically-based
conceptual framework for quantum gravity. A few years later, a specific 
aspect of those ideas was considered by Viola and  Onofrio \cite{vo}; and 
subsequently, in greater detail, by Delgado \cite{vd}. Concurrently, one of us 
has pursued this subject in the context of neutrinos, and for other 
systems modeled after neutrinos \cite{ws1,grf1997}, while
several other authors have made significant progress in understanding
these effects  in the context of flavor oscillation clocks and Brans-Dicke 
theory \cite{ws3}. Independently, in early 1999, Amelino-Camelia was to 
falsify the impression that quantum-gravity induced space-time fluctuations
\cite{jaw} are beyond the scope of terrestrial probes. The  Amelino-Camelia 
argument is fast becoming a classic and the reader is
referred to Refs. \cite{gacN} for its origins. The astrophysical neutrinos, 
and gamma ray bursts, are also being viewed as probes of certain 
quantum-gravity aspects of space-time \cite{grb1}. The general point to made, 
therefore, is that experimental and observational techniques have reached 
a point where various quantum aspects of gravity can now be probed 
terrestrially. The results reported here and those found in \cite{gacN}, 
if confirmed, could lead to first experimental signatures indicating a 
profound difference between the classical and quantum aspects of gravity.

The abstracted thesis lies in the observations: 

\begin{enumerate}
\item[A.]
In contrast to the simplest local $U(1)$
gauge theory coupled to electrically charged
spin-$1/2$ matter, i.e. theory of  quantum electrodynamics, quantum gravity 
is {\em not\/} endowed with a super-selection rule
prohibiting a linear superposition of different gravitational charges.

\item[B.]
For quantum objects in a linear 
superposition of different mass (or, energy) eigenstates, there is an 
operational limit on the fractional accuracy beyond which quantum measurement 
theory forbids any claim on the equality of the inertial and gravitational 
masses. 
\end{enumerate}

Since we wish to stay as close to the terrestrial experiments as possible we
present our arguments in the context of the weak gravitational 
fields. In particular, in order that a possible violation of the
equivalence principle can be studied without theoretical prejudices
we will treat gravitational interaction on the same footing as 
any other interaction. This means that any changes in the clocks and
rods will be manifestly dependent on the gravitational environment.
In the event there is no violation of the equivalence principle, the
affect on rods and clocks would reduce to the expectations based
on general relativity.

\section{Weak-field limit of quantum gravity and absence of a 
super-selection rule}

The 1975 experiment of Colella, Overhauser, and Werner (COW) 
on the gravitationally induced phases in neutron interferometry
established the weak-field limit of quantum  gravity 
for the 
non-relativistic regime to be 
\cite{COW1975}\footnote{In the context of this equation, a knowledgeable 
physicist wrote to one of us (DVA)
 `` ... quantum gravity is the theory of the quantum properties
of the gravitational field, not the study of the properties of
quantum matter in a gravitational field.  You work in a different
context than quantum gravity: the context of quantum matter 
interacting with classical gravity. Isn't it?'' Yes, and yet if 
experiments found that quantum matter interacting with classical
gravity carries unexpected features, then the envisaged theory
of quantum gravity cannot remain immune to such a development. It is
in this realm of ``quantum gravity'' that this {\em Letter\/} is set.} 
\eqb
\left[-\left(\frac{\hbar^2}{2\,m_i}\right)
\nabla^2+ m_g c^2\, \Phi\right]
\psi(\vec x,t)= i\hbar\frac{\partial\psi(\vec x,t)}{\partial t}
\label{sheq}
\eqe
where $\Phi$ is the dimensionless gravitational potential.
In the kinetic term, $m_i$  is the inertial 
mass of a particle, and in the interaction term, 
 $m_g$ is its gravitational mass. To be precise, the 
1975 {\sc COW}-experiment established this equation
for neutrons and verified the equality, $m_i=m_g$ for neutrons, to 
an accuracy of roughly
$1\%$. Since then these measurements have become more refined.
Among the most notable confirmations of this result are by a recent 
atomic-interferometry experiment at Stanford 
\cite{chuetal}, and by an experiment at Institute Laue-Langevin
using a new type of neutron interferometer \cite{gvdz}.
The Stanford experiment  has established that a macroscopic 
glass object falls with the same gravitationally-induced acceleration,
to within $7$ parts in $10^9$, as a cesium atom in linear superposition
of two different energy eigenstates. 
The experiment at Institute Laue-Langevin
indicates that a statistically significant anomaly observed in the silicon
single-crystal interferometer \cite{COW1997} can be excluded by 
more than one standard-deviation. Clearly, one eagerly awaits
more data from  Institute Laue-Langevin to unambiguously
interpret the anomaly reported in Ref. \cite{COW1997}.
If the theoretical framework presented in this {\em Letter\/} 
turns out to be correct, we should expect 
the Stanford experiment with a significant improvement
in its already-impressive accuracy to see a difference in the
gravitationally-induced accelerations of the classical and quantum
objects.

\subsection{Terrestrial Gravitational Environment}

From the perspective of terrestrial
experiments, $\Phi$ contains two important contributions: $\Phi(z)$, 
and $\widehat\Phi$. The former has Earth as its source, and the
latter arises from the cosmic distribution of matter. As a study of the
planetary motions reveals, and as simple calculations confirm, 
$\widehat\Phi$ is essentially constant over the dimensions of
the solar system. While in the context of the
theory of general relativity it is difficult
for local observations to detect  
$\widehat\Phi$, a  $\widehat\Phi$  acquires local observability
through a violation of the equivalence principle. A violation that,
we shall show, is inherent to quantum gravity. 

Explicitly, under the boundary condition that $\Phi(\infty)$
vanishes at spatial infinity, $\Phi(z)$ is given by:
\eqb
\Phi(z)\approx -7\times 10^{-10} \frac{R_\oplus}{R_\oplus+z}
\eqe
where $R_\oplus$ is radius of the earth, and $z$ is the vertical distance 
from the surface of the Earth. The remark on  $\Phi(\infty)$ also implies that
we shall restrict to the ``local quasi-Newtonian'' neighborhood (LQNN)
\eqb
{\sc LQNN}:\quad R_{Milky\,\, Way}\ll R_{\sc LQNN} \ll R_{Hubble}
\eqe
where $R$ refers to dimensions of the indicated region.
It is not surprising that an estimate of $\widehat\Phi$ is a more involved
procedure.
Towards that end we note that
a compilation of roughly three hundred galaxy clusters shows that the 
super-clusters they define  are in a lattice-like 
distribution with cells
430 million light-years in size \cite{je,md}. The whole distribution
resembles a ``three-dimensional chessboard'' \cite{je,rbt}. It is
the gravitational environment created by such a structure that interests
us.  For laboratory regions
much smaller than the lattice size, 
a phenomenological description of the
gravitational environment contains,  (a)
a potential gradient that is annulled (for an observer at rest
inside the laboratory) 
by the gravitationally induced acceleration of the laboratory
under consideration,  and (b) a non-acceleration inducing constant 
gravitational potential. To see this clearly, imagine all the 
${\sc LQNN}$-matter
uniformly distributed around the region of interest with a spherical
symmetry. In this approximation, there are no gradients in the gravitational
potential and  the effect of the local cosmic distribution of matter can be
phenomenologically replaced by a constant gravitational potential.
It is also to be noted that this still assumes, $\Phi(\infty)=0$, and 
only those contributions are included in $\widehat\Phi$ 
for which the Newtonian 
condition $\Phi\ll 1$ is still valid.
To obtain an estimate for the  contributions to $\widehat\Phi$, we
note that a local super-cluster known as Shapley Concentration (SC) has
a mass around $10^{16}$ solar masses, 
and it is at a distance of about 
$700$ million light-years from us. This information yields:     
\eqb
\widehat\Phi_{\sc SC} \approx -\,2 \times 10^{-6}
\eqe
Similarly, the Great Attractor's (GA)  contribution to $\widehat\Phi$  
is \cite{ik}:
\eqb
\widehat\Phi_{\sc GA} \approx -3\,\times 10^{-5}
\eqe
In the presence of a non-zero cosmological constant these considerations
can be modified in a well defined manner \cite{mn}.

In the above scenario 
\eqb
\vert\widehat\Phi\vert =  \vert\widehat\Phi_{\sc GA}\vert + 
\vert\widehat\Phi_{\sc SC}\vert +
\cdots \gsim 3.2\times 10^{-5}\label{zzz}
\eqe
where the unspecified contributions contain 
the influence of the entire LQNN.
The total $\widehat\Phi$ must necessarily involve general-relativistically 
defined and calculated contributions from the entire cosmic matter. 
The purpose of expression (\ref{zzz}) is only to serve as 
an existence argument for, and a lower bound on, $\vert\widehat\Phi\vert$.  
It is to be noted that 
\eqb
{\vert\widehat\Phi\vert}/{\left\vert\Phi(z=0)\right\vert} 
\gsim 0.5\times 10^{5}\label{gga}
\eqe
Despite the fact that $\vert\widehat\Phi\vert$ exceeds 
${\left\vert\Phi(z=0)\right\vert}$ by five orders of magnitude,
the observability of $\Phi(z)$, as seen for example through the lunar orbit,
arises because
\eqb
\left\vert \vec\nabla\Phi(z)\right\vert \gg \left
\vert\vec\nabla\widehat\Phi\right\vert \label{ggb}
\eqe
The insensitivity of the planetary orbits to
$\widehat\Phi$ resides in its
essential  constancy over the solar system and in 
the validity of the equivalence principle for classical objects to
a high precision \cite{tp1,tp2}. 
The $\widehat\Phi$ does contribute to galactic motion.
 
In the context of the solar neutrino anomaly, and a violation
of the equivalence principle,  we shall show that
$\widehat\Phi$ acquires a local observability. In general, 
in any quantum gravity phenomenology which does not
{\em a priori\/} exclude a violation of the equivalence principle (VEP) as
a possibility, 
$\widehat\Phi$ can carry significant physical consequences.
In fact, we shall show that if
the inertial and gravitational masses are considered as operationally
distinct objects, quantum gravity carries an inherent
quantum induced violation of the equivalence principle (qVEP).

\subsection{Flavor-oscillation clocks as probes 
of qVEP and $\widehat\Phi$} 

Having once defined the gravitational environment of interest, 
we now remind the reader that the standard text-book  \cite{jjs}
understanding of the neutron interferometry is based on the
Schr\"odinger equation (\ref{sheq}) 
with $\Phi$ replaced by $\Phi(z)$.
Such a  treatment  completely ignores all possible effects 
due to $\widehat\Phi$.
This is entirely justified for a single mass eigenstate in a 
non-interferometry setting (and without a VEP).
However,  for states that are in linear 
superposition of different masses, or, more generally,  energies, 
a quantum-gravity phenomenology that wishes to study any possible
violations of the equivalence principle
cannot {\em a priori\/} ignore  $\widehat\Phi$ \cite{mg1,mg2}.

In particular, if one allows  for a violation of the equivalence principle,
under  the correction\footnote{This does not 
constitute a transformation of changing $\Phi(\infty)$ by a constant.
Both $\Phi(z)$ and $\widehat\Phi$ have been calculated 
with $\Phi(\infty)=0$.}
\eqb
\Phi(z) \rightarrow \Phi(z) + \widehat\Phi,
\eqe
equation (\ref{sheq}) is not invariant for
states that are in a linear superposition of different energy, 
or different mass, eigenstates.
The argument that establishes this is simple, but non trivial. Therefore, 
the reader's attention is invited to the details.
We construct two flavor states
at a time $t=0$\footnote{See remarks contained in the last paragraph 
at the end of Sec. 1.}
\beq
\vert \alpha\rangle &=& \,\,\,\,\,\cos\theta\,\vert m_1\rangle + 
\sin\theta\,\vert m_2\rangle\nonumber\\ 
\vert \beta\rangle &=& -\sin\theta\,\vert m_1\rangle + 
\cos\theta\vert\, m_2\rangle\label{flavor}
\eeq
Let each of the mass eigenstates, $\vert m_1\rangle$
and  $\vert m_2\rangle$, be at rest. The idealized frame
of observations is chosen to be such that the only gravitational potential
present is $\widehat\Phi$.
Then, the probability amplitude for a flavor oscillation from the state  
$\vert \alpha\rangle $
to the state $\vert \beta\rangle$ at a later time, $t>0$, is
\beq
{A}_{\alpha\rightarrow\beta}(t)=
&&\langle \beta\vert{\Bigg\{}\cos\theta\exp\left[ -i \frac{\left(
m^i_1+m^g_1\widehat\Phi\right) c^2 \,t}
{\hbar}\right]
\vert m_1\rangle\nonumber\\
&&\qquad\qquad+
\sin\theta\exp
\left[
 -i \frac{\left(
m^i_2+m^g_2\widehat\Phi\right) c^2\, t}
{\hbar}\right]
\vert m_2\rangle {\Bigg\}}\label{a}
\eeq
The 
superscripts ``i'' and ``g'' on ``$m$''  
refer respectively to inertial and gravitational masses.

We draw attention to the fact that each
of the mass eigenstates picks up a different, mass-dependent, 
gravitationally induced phase from $\widehat\Phi$.
In consequence, these relative phases
become experimentally observable.
This observability is in the usual sense in which one measures red shifts
as, e.g., done by a comparison of clocks located in regions characterized by
a different $\widehat\Phi$.
However, if the principle of equivalence is violated, then  $\widehat\Phi$
acquires a local observability. It is precisely the absence of a 
super-selection prohibiting the flavor states, simplest of which are
given by $\vert \alpha\rangle $ and $\vert \beta\rangle$, that allows
for emergence of the $\widehat\Phi$-dependent relative phases. 
Had a super-selection rule prohibited the existence of 
flavor states, i.e. demanded
$\theta=0$, then the emergent $\widehat\Phi$-dependent
phase would have been a global one, and thus it would
have carried no physical 
observability.

Incidently, as it emerges from the above discussion,
a freely falling frame, contrary to
the usual assertions, is not necessarily a frame devoid of a 
gravitational field. Vanishing of the curvature tensor is a necessary,
but not sufficient, condition
for the absence of gravity.  

The probability amplitude (\ref{a}) yields
probability for the flavor oscillation from the state  
$\vert \alpha\rangle $
to the state $\vert \beta\rangle$:
\beq
{ P}_{\alpha\rightarrow\beta}(t) =  { A}^\ast_{\alpha\rightarrow\beta}
(t)
 \,{ A}_{\alpha\rightarrow\beta}(t) =
\sin^2\left(2\theta\right)\sin^2\left(\omega_{\alpha\rightarrow\beta}
\,t\right)\label{fo}
\eeq
where the angular frequency for the flavor 
oscillations is,
\beq
\omega_{\alpha\rightarrow \beta}=
 \frac{c^2}{2\hbar}\left(m^i_2-m^i_1\right)+
 \frac{c^2}{2\hbar}\left(m^g_2\widehat\Phi
-m^g_1\widehat\Phi\right)\label{w}
\eeq
The flavor-oscillation frequency, $\omega_{\alpha\rightarrow \beta}$,
consists of two parts. The first is the kinetic part, and the second
is the  $\widehat\Phi$ contribution. In a distant  
planetary system,
where the local gravitational environment can be significantly different,
the second part would be different. The existence 
of time-periodicity via $\omega_{\alpha\rightarrow \beta}$
allows us to interpret the system of flavor states  
as a clock. The flavor-oscillation clocks red-shift via the 
$\widehat\Phi$-contribution to 
$\omega_{\alpha\rightarrow \beta}$. On {\em assuming\/} the 
equality of the inertial and gravitational masses in this setting,
the general-relativistic predicted red shift is reproduced. This
can be measured by comparing this clock with another embedded in a 
distant region with a different $\widehat\Phi$. So far all we have established
is that red-shift of flavor oscillations clocks shows a deep consistency in
the quantum mechanical evolution and the principle of equivalence as embedded
in the equality of the inertial and gravitational masses.

We now turn to a fundamentally quantum mechanical source of a VEP. 
In what follows we shall take the view that  
the inertial and gravitational masses are two independent objects, a view
compatible with Lyre's recent critique of the 
subject \cite{hl}. This view has the further support in that
the very operational procedures that define inertial and
gravitational masses are profoundly different, and it is more so in the 
quantum context.

The essential physical idea is that if any one of the two 
quantities $A$ and $B$
carries an intrinsic quantum uncertainty, $\Delta$, then from an operational
point of view the equality
of $A$ and $B$ cannot be claimed beyond the fractional accuracy 
\beq
\frac{\Delta}{(A+B)/2}
\eeq 
We shall consider this general statement in the context of the equality
of the inertial and gravitational masses to establish a quantum-induced
violation of the equivalence principle (qVEP):
\beq
m_g=\left(1+f\right) m_i
\eeq
where $f$ vanishes for objects with a classical counterpart and is
non-zero for objects with no classical counterpart.
Coupled with  result (\ref{w}), this  implies that
flavor oscillation clocks carry a local observability for
$\widehat\Phi$. That is, while in the absence of a qVEP 
(or, even in the absence of a  VEP) $\widehat\Phi$ has no local observability. 
A qVEP/VEP can make a $\widehat\Phi$ observable locally. This is an important
implication of the violation of equivalence principle. Quantum systems are
particularly sensitive to a $\widehat\Phi$ as can be seen by  
referring to Eqs. (\ref{gga}) and (\ref{ggb}), 
and on taking note of the observations: (a) 
Classically important gravitationally-induced force in being
proportional to $\vec\nabla\widehat\Phi$ carries relatively little
physical significance in the present context, and
(b) Quantum mechanically important gravitationally-induced relative phases
are proportional,
apart from the difference in masses of the underlying mass eigenstates,
to $f\widehat\Phi$.

We now immediately note that the
very quantum construct that defines the flavor eigenstates does not allow
them to carry a definite mass. Therefore, within the stated framework, the
equality of the inertial and gravitational masses
looses any operational meaning
beyond the flavor-dependent fractional accuracy defined as
\beq
f_\eta\equiv\frac{\sqrt{\langle\nu_\eta\vert \widehat m^2\vert\nu_\eta\rangle
-
\langle\nu_\eta\vert \widehat m\vert\nu_\eta\rangle^2}}
{\langle\nu_\eta\vert \widehat m\vert\nu_\eta\rangle},\quad\eta=\alpha,\beta
\eeq
Here $\widehat m$ is the mass operator: $\widehat m\vert m_\jmath\rangle
= m_\jmath \vert m_\jmath\rangle$. 
Explicitly, this yields:
\beq
f_\alpha=\frac{\sin(2\theta) \delta m}
{2\left(m_1+\sin^2(\theta)\delta m\right)},\quad
f_\beta=\frac{\sin(2\theta) \delta m}
{2\left(m_1+\cos^2(\theta)\delta m\right)}
\eeq
where $\delta m\equiv m_2 - m_1$. For $\delta m \ll m_\imath, \quad\imath
=1,2$, we  have the flavor-independent relation
\eqb
f=\frac{\sin(2\theta)\delta m}{2 \langle m\rangle}\label{f}
\eqe
Only for some very specific values of $\theta$, i.e. for $2 \theta=n\,\pi,
\quad n=0,1,2,\cdots$, 
do the states $\vert\alpha\rangle$ and  $\vert\beta\rangle$ coincide
with states characterized by  well-defined masses. For such states 
alone $f\,$ vanishes. Otherwise, $f\,$ remains non-vanishing.

Once the equality of the inertial and gravitational masses has been 
compromised,
the potential $\widehat\Phi$ carries not only observability only in
comparison with ``distant'' systems  but it also becomes observable 
locally.

\section{Illustrative Experimental Settings}

We now explore how the above obtained qVEP
can be studied in some realistic experimental settings.

\subsection{Atomic interferometry}

Consider two sets, differentiated by the index ``$\imath$,''  
of ``flavors'' for Cesium atoms:
\beq
\left[
\matrix{\vert ^{\alpha}Ce\rangle_{\xi_\imath}\cr
        \vert ^{\beta}Ce\rangle_{\xi_\imath}}\right]
= \left[\matrix{\,\,\,\,\,\cos(\xi_\imath) & \sin(\xi_\imath) \cr
	-\sin(\xi_\imath) & \cos(\xi_\imath)}\right]
\,
\left[
\matrix{\vert ^{E_1}Ce\rangle\cr
        \vert ^{E_2}Ce\rangle}\right],\quad \imath=a,b
\eeq 
Here, $\vert ^{E_1}Ce\rangle$ and $\vert ^{E_2}Ce\rangle$ represent two
different {\em energy\/} eigenstates of the Cesium atom. 
The ``flavor'' states, $\vert ^{\alpha}Ce\rangle_{\xi_\imath}$ and 
$\vert ^{\beta}Ce\rangle_{\xi_\imath}$, 
are linear superposition of the energy eigenstates and  
are characterized by the flavor indices $\{\alpha, \beta\}$,
and by the mixing angle $\xi_\imath$. That is, we have two copies 
of the flavor states similar to the ones introduced in Eqs. (\ref{flavor}).
Each of the two copies is defined by a {\em different\/} mixing angle.

As observed in identical  free fall experiments by a stationary
observer on Earth, the flavor-dependent qVEP predicts a {\em fractional 
difference in the
spread in their accelerations\/} to be:
\beq
 \frac{\vert \Delta a_{\ell\xi_b}\vert
 -  
\vert \Delta a_{\ell\xi_a}\vert}{g} \approx
\left\{\left\vert
\frac{\sin(2\xi_b)}{\langle E_{\ell\xi_b}\rangle}
\right\vert
-
\left\vert
\frac{\sin(2\xi_a)}
{\langle E_{\ell\xi_a}\rangle} \right\vert
\right\}
\frac{\delta E}{2},\quad\ell=\alpha,\beta
\eeq
where $\delta E \equiv E_2-E_1$, and
$g$ refers to the magnitude of the acceleration due
to gravity for a classical object. 
One may for instance take $\xi_a=0$, and
$\xi_b=\pi/4$. Then the above expression reduces to
\eqb
 \frac{\vert \Delta a_{\xi_b=\pi/4}\vert}{g}
 \approx
\frac{\delta E}{2\,{\langle E_{\xi_b}\rangle}}
\eqe
where the $\ell$ dependence has dropped because to the lowest 
order $f_\ell$ is {\em same\/} for both flavors [{\em not\/} both sets, see 
Eq. (\ref{f})].
For flavor states of Cesium atoms prepared with $\delta E 
\approx 1 \,\,\mbox{eV}$, this difference is of the order of a few parts
in $10^{12}$, and should be observable in refined versions of experiment
reported in Ref. \cite{chuetal}. How difficult this refinement in 
techniques at Stanford and NIST would be is not fully known to us.
However, the extraordinary accuracy in similar experiments and the 
already achieved absolute uncertainty of $\Delta g/g\approx 3\times 10^{-9}$,
representing a million fold increase compared with previous experiments,
makes us cautiously optimistic about observing qVEP in atomic 
interferometry experiments pioneered  by the group of Steven Chu at Stanford.

At the same time, a  definitive null result for qVEP would establish that the 
inertial and gravitational mass, despite their 
different operational definitions, are physically identical objects.

\subsection{Solar Neutrino Anomaly}

In recent times it has
been conjectured that the solar neutrino anomaly may be related
to a flavor-dependent violation of the equivalence principle
\cite{solarVEP1,srp2,z1,z2}.
The suggestion, in fact,  originally came from Maurizio Gasperini 
in the late 1980's \cite{mg1,mg2}.
After appropriate generalization, 
and on interpreting
the flavor index $\ell$ to represent the three neutrino flavors ($\nu_e$,
$\nu_\mu$ and $\nu_\tau$),
the non-relativistic arguments presented so far can be readily 
extended to the relativistic system of neutrino oscillations.
Here we simply provide an outline of this argument.

To estimate the qVEP effects it suffices to restrict
to a two-state neutrino oscillation framework. A simple calculation shows
that the difference in fractional measure of qVEP turns out to be 
exceedingly small \cite{grqc0002005}:
\beq
\Delta f_{\ell\ell'}\equiv f_\ell -f_{\ell'}  = 6.25 \times 10^{-26} \left[
\frac{(\Delta m^2)^2}{\mbox{eV}^4}\right]
\left[\frac{\mbox{MeV}^4}{E^4}\right] \sin(4\theta_V)  
\eeq
where $\ell$ (say, $\nu_e$) 
and $\ell'$ (say, $\nu_\mu$) refer to two different neutrino flavors, and
$\theta_V$ is the vacuum mixing angle between the underlying mass
eigenstates (whose superposition leads to different flavors of neutrinos).
The difference in the squares of the underlying mass eigenstates, 
$m_2^2-m_1^2$, has been represented  by $\Delta m^2$ (in eV$^2$);  
and $E$ is the 
expectation value of the neutrino energy (in MeV). 
The qVEP-induced oscillation length is found to be \cite{grqc0002005}
\beq
\lambda^{osc}_{qVEP} = \left[0.66\times 10^2\,
\frac{E^3}{\left(\Delta m^2\right)^2
\vert \sin(4 \theta_V)\widehat\Phi\vert}\right]
\times \lambda_\odot\label{L}
\eeq 
where $\lambda_\odot \simeq 1.5\times 10^{8}\,\,\mbox{km}$
is the mean Earth-Sun distance.

Assuming that the mass and gravitational eigenstates coincide,
the oscillation length that enters the 
neutrino-oscillation probability is not $\lambda_{qVEP}^{osc}$
but  \cite{mg2}
\beq
\lambda^{osc} = \frac{\lambda_0^{osc} \lambda_{qVEP}^{osc}}
{\lambda_0^{osc} +  \lambda_{qVEP}^{osc}}
\eeq
where $\lambda_0^{osc}$ is the usual 
kinematically induced oscillations length
\beq
\lambda^{osc}_0= \left[\frac{2\pi\,\, \mbox{m}}{1.27}\right]
\left[\frac{E}{\mbox{MeV}}\right]
\left[\frac{\mbox{eV}^2}{\Delta m^2}\right]
\eeq
 For $\Delta m^2$, the LSND
excess event anomaly \cite{LSND} 
sets $\Delta m^2 \simeq 0.4 \,\,\mbox{eV}^2$, while the Super-Kamiokande
\cite{SuperK} 
evidence on atmospheric neutrino oscillations suggests another
 $\Delta m^2 \simeq 3\times 10^{-3} \,\,\mbox{eV}^2$. For both of these
mass-squared differences, for any set of
reasonable parameters,  $\lambda_{qVEP}^{osc}
\gg \lambda_{0}^{osc}$. Consequently,  $\lambda^{osc}\simeq
 \lambda_{0}^{osc}$, and the qVEP-induced effects is
suppressed by the kinematic term.  

On the other hand, there is yet no independent confirmation for the
LSND result, and the possibility remains open as  to whether
two of the three underlying mass
eigenstates are massless. In that event, the relevant oscillation length
associated with  qVEP is:
\beq
\lambda_{qVEP}^{osc} = \frac{2\pi c \hbar}{\widehat\Phi
\left(\zeta_{\ell\ell^\prime}/E\right)E} =
\frac{2\pi c \hbar}{
\zeta_{\ell\ell^\prime}\,\,\widehat{\Phi}}
\simeq
\left(\frac{\pi}{\zeta_{\ell\ell^\prime}\,\widehat\Phi}\right)
 \times 10^{-33}
\,\,\mbox{km}
\eeq
where, in the {\em last\/}
term,  $\zeta_{\ell\ell^\prime}$ is measured in $eV$, and is defined as: 
\beq
\zeta_{\ell\ell^\prime} \equiv\left\vert \sqrt{
\langle \nu_\ell\vert  H^2\vert \nu_\ell\rangle
- \langle \nu_\ell\vert  H\vert \nu_\ell\rangle^2}
-
\sqrt{
\langle \nu_{\ell^\prime}\vert  H^2\vert \nu_{\ell^\prime}\rangle
- \langle \nu_{\ell^\prime}\vert  H\vert \nu_{\ell^\prime}\rangle^2}
\right\vert
\eeq
This oscillation length has no explicit energy dependence, except
that carried via  $\zeta_{\ell\ell^\prime}$. The  $\zeta_{\ell\ell^\prime}$
measures the difference in the energy-widths of the
wave functions of the the two involved flavors.

If this was to be a solution to the solar neutrino anomaly, it requires that
$\lambda^{osc}_{qVEP} \sim \lambda_\odot$, i.e.
\eqb  
{\zeta_{\ell\ell^\prime}\,\widehat\Phi}\sim 2\times 10^{-41}\,\,\mbox{eV}
\eqe
This requirement
tells us  that the wave functions of the two involved flavors carry the
same width to a very high precision.

This solution to the solar neutrino anomaly requires a non-zero
$\widehat\Phi$, and shows this to be locally observable.

In terms of the $E^n$ dependence,
the kinematic oscillation length carries $n=1$, the Gasperini-conjectured
VEP for massless neutrinos 
has $n=-1$, and the qVEP-induced oscillation length for the massless
neutrinos has $n=0$ ({\em cf.\/} n=3, for massive qVEP).

Therefore, as more data on the solar neutrino flux
becomes available we should be able to distinguish between the 
various mechanisms.
Of course, a possibility that more than one mechanism is at play in reality
should not be ignored.

\section{Concluding Remarks}

Several new  theoretical insights, and an outline for new experimental
proposals, emerge in this {\em Letter.\/} 
First, we noted that
the super-selection rule that prohibits the linear superposition of
states with different electric charges, has no counterpart in the
realm of gravitational interactions. The absence of this super-selection
rule endows the cosmic gravitational potentials 
in quantum gravity with a much more visible physical status,
without altering the classical results. 
Under the assumption that the inertial and 
gravitational masses are operationally different objects, we showed that since
flavor states carry an inherent quantum uncertainty in their masses (or
energies)   the equality of the their inertial and gravitational 
masses looses operational meaning beyond certain fractional accuracy.
We used this fact to make a prediction on the
free fall of states that are in a linear superposition of different 
energy eigenstates. That prediction can be tested in the new generation
of atomic interferometry experiments. Furthermore, we established that
a qVEP can also have significant implications for the solar neutrino
anomaly. 
Elsewhere we shall sketch how the Schr\"odinger's 
SQUID \cite{squids}., 
i.e. a superconducting quantum interference device with a linear 
superposition of macroscopic counter-propagating supercurrents, could serve as 
a sensitive probe of Earth's gravitomagnetic field.

{\bf Acknowledgements}

We are indebted to a referee for asking us many questions and 
for making several suggestions to improve the presentation of this 
{\em Letter.\/} These questions
are attended here by for several pedagogic remarks. Maurizio Gasperini, 
 Roland Koberle, and
Carlo Rovelli, are thanked for their insightful
questions and ensuing discussions. I (DVA) extend my thanks to 
Hugo Morales for  bringing their work, cited in Ref. \cite{grb1}, to my 
attention, and for a discussion on the details of that paper. 
It is also my (DVA) pleasure to thank Sam 
Werner for bringing several aspects of neutron interferometry 
to my attention over the last few years. I thank G. van der Zouw for
providing reference \cite{gvdz}, and for a valuable correspondence.




\begin{thebibliography}{000}


\bibitem{gacN}
G. Amelino-Camelia,  Nature  398 (1999) 216;
G. Amelino-Camelia, Phys. Lett B 477 (2000) 436;
G. Amelino-Camelia, gr-qc/9910089; gr-qc/9910023; gr-qc/9903080;
Y. J. Ng,  H. van Dam, Phys. Lett. B 477 (2000) 429.

\bibitem{dvaN}
D. V. Ahluwalia, Nature  398 (1999) 199.

\bibitem{ahl} 
J. Audretsch, F. W. Hehl, C. L\"ammerzahl, 
``WE-Haraeus Seminar 1991:0368-408,'' edited by J. Ehlers, G. Sch\"afer
(Springer, Berlin, 1991). Also see, T. Goldman, R. J. Hughes, M. M. Nieto,
Phys. Lett. B 171 (1986) 217. 

\bibitem{vo}
L. Viola, R. Onofrio,
Phys. Rev. D  55 (1997) 455;
R. Onofrio, L. Viola, Mod. Phys. Lett. A  12 (1997) 1411.

\bibitem{vd}
V. Delgado, Phys. Rev. A  59 (1999) 1010.


\bibitem{ws1}
D. V. Ahluwalia,  C. Burgard, 
Gen. Rel. Grav.  28 
(1996) 1161; (E)  29 (1997) 681;
D. V. Ahluwalia, and C. Burgard,
Phys. Rev. D  57 (1998) 4724.

\bibitem{grf1997}
D. V. Ahluwalia, 
Gen. Rel. Grav.  29 (1997) 1491 -- an interested reader should make 
amend for some obvious ``cut-and-paste'' wrong placings of the paragraphs
on publisher's end. {\em Cf.} 
B. Mashhoon, gr-qc/0003022.


\bibitem{ws3}
K. Konno,  M. Kasai,
Prog. Theor. Phys.  100 (1998) 1145;
Y. Grossman,  H. J. Lipkin, 
Phys. Rev. D  55 (1997) 2760;
A. Camacho, Mod. Phys. Lett. A  14 (1999) 2245;
S. Capozzielo, G. Lambiase,  Mod. Phys. Lett. A  14 (1999) 2193.

\bibitem{jaw}
J. A. Wheeler, in {\em Relativity, Groups and Topology\/}, edited by
B. S. DeWitt and C. M. DeWitt (Gordon and Breach, New York, 1964);
L. J. Garay, Phys. Rev. Lett.  80 (1998) 2508;
S. Carlip,  Phys. Rev. Lett.  79 (1997) 4071;
R. Garattini,  Phys. Lett. B  459 (1999) 461.

\bibitem{grb1}
J. Ellis, N. E. Mavromatos,  D. V. Nanopoulos,
Phys. Rev. D  61 (2000) 027503;
G. Amelino-Camelia, J. Ellis,  N. E. Mavromatos,  D. V. Nanopoulos,
S. Sarkar, Nature  393 (1998) 763;
S. D. Biller {\em et al.}
Phys. Rev. Lett. 83 (1999) 2108;
J. Alfaro. H. A. Morales-T\'ecotl,  L. F. Urrutia,
{\em Phys. Rev. Lett.\/} 84 (2000) 2318. In the 
paper by Alfaro {\em et al.} quantum gravity
effects are shown to induce a time delay for neutrinos that
have traveled cosmological distances. Curiously,
the dominant contribution is helicity {\bf in}dependent. 
Whereas for photons, a similar calculation yields a 
helicity dependent effect \cite{gp}. 
It appears to one of us (DVA) that these apparently different helicity 
dependences arise from lack of realization by the authors of this work
that Majorana spinors form a bi-orthonormal set of spinors and that
their usual norm is identically zero (even for massive particles) while
their ``bi-ortho norm''
is imaginary definite. These aspects of Majorana spinors
can be found in ref. \cite{dva_maj}.

\bibitem{COW1975}
R. Colella, A. W. Overhauser, S. A. Werner,
Phys. Rev. Lett. 34 (1975) 1472.


\bibitem{chuetal}
A. Peters, K. Y. Chung, S. Chu. Nature  400  
(1999) 849.

\bibitem{gvdz}
G. van der Zouw, {\em et al.\/}, Nucl. Instr. and Meth. Phys. Res.
440 (2000) 568.

\bibitem{COW1997}
K. C. Littrel, B. E. Allman,  S. A. Werner,
Phys. Rev. A  56 (1997) 1767.

\bibitem{je}
J. Einasto, Nature  385  (1997) 139.

\bibitem{md}
M. Drinkwater, Science 287 (2000) 1217.

\bibitem{rbt}
R. B. Tully {\em et al.\/} Ap. J. 388  (1992) 9.

\bibitem{ik}
I. R. Kenyon, Phys. Lett. B  237 (1990) 274.

\bibitem{mn}
M. Nowakowski, gr-qc/0004037.


\bibitem{tp1}
G. L. Smith {\em et al.\/}, 
Phys. Rev. D  61 (2000)  022001.

\bibitem{tp2}
S. Bae$\beta$ler, {\em et al.},
Phys. Rev. Lett.  83 (1999) 3585.

\bibitem{jjs}
J. J. Sakurai, {\em Modern Quantum Mechanics\/} (Benjamin/Cummings 
Publishing Co., California, 1985). 


\bibitem{mg1}
M. Gasperini, Phys. Rev. D 38 (1988) 2635.

\bibitem{mg2}
M. Gasperini,
Phys. Rev. D 39 (1989) 3606.


\bibitem{hl}
H. Lyre, gr-qc/0004054


\bibitem{solarVEP1}
J. N. Bahcall, P. I. Krastev, C. N. Leung, Phys. Rev. D  52
(1995) 1770;
A. Halprin, C. N. Leung, J. Pantaleone, 
Phys. Rev. D.  53  (1996) 5365;
J. R. Mureika,
Phys. Rev. D  56 (1997) 2408;
A. M. Gago, H. Nunokawa, R. Zukanovich Funchal, hep-ph/9909250;
S. W. Mansour,  T. K. Kuo,
Phys. Rev. D  60 (1999)  097301.


\bibitem{srp2}
G. L. Fogli, E. Lisi, A. Marrone,  G. Scioscia,
Phys. Rev. D. 60 (1999)  053006;
E. Lisi, A. Marrone, and D. Mantanino,  hep-ph/0002053. 

\bibitem{z1}
D. V. Ahluwalia, Mod. Phys. Lett. A 13 (1998) 2249.

\bibitem{z2}
D. V. Ahluwalia, Mod. Phys. Lett. A 13 (1998) 2249;
I. Stancu, D. V. Ahluwalia, Phys. Lett. B. 460 (1999) 431;
V. Barger, S. Pakvasa, T. J. Weiler, K. Whisnant, Phys. Lett. B
437 (1998) 107; H. Georgi, S. Glashow, Phys. Rev. D 61
(2000) 097301; A. J. Baltz, A. S. Goldhaber, M. Goldhaber,
Phys. Rev. Lett. 81 (1998) 5730.

\bibitem{grqc0002005}
D. V. Ahluwalia, gr-qc/0002005.

\bibitem{LSND}
C. Athanassopoulos, {\em et al.}, Phys. Rev. Lett. 81 (1998) 1774.

\bibitem{SuperK}
Y. Fukuda {\em et al.}, Phys. Rev. Lett. 81 (1998) 1562.

\bibitem{gp}
R. Gambini, J. Pullin,  Phys. Rev. D 59 124021;

\bibitem{dva_maj}
D. V. Ahluwalia, Int. J. Mod. Phys. A 11 (1996) 1855.  


\bibitem{squids}
A. Cho, Science 287 (2000) 2395.


\end{thebibliography}
\end{document}